\documentclass[paper]{ieice}
\usepackage{graphicx,xcolor}
\usepackage[fleqn]{amsmath}
\DeclareMathOperator*{\argmin}{arg\,min}

\usepackage{bm}
\usepackage{subfigure}
\usepackage{cite}
\usepackage{comment}
\usepackage{enumitem}

\usepackage{color}

\field{A}
\title{Hue-Correction Scheme Considering Non-Linear Camera Response for Multi-Exposure Image Fusion}
\authorlist{%
 \authorentry{Kouki SEO}{n}{A}\MembershipNumber{}
 \authorentry{Chihiro GO}{n}{A}\MembershipNumber{}
 \authorentry{Yuma KINOSHITA}{s}{A}\MembershipNumber{1614118}
 \authorentry[kiya@tmu.ac.jp]{Hitoshi KIYA}{f}{A}\MembershipNumber{7904021}
}
\affiliate[A]{The authors are with Department of Computer Science, Tokyo Metropolitan University, Hino-shi, 191-0065 Japan.
}

\received{2015}{1}{1}
\revised{2015}{1}{1}

\begin{document}
\setlength{\textfloatsep}{7pt}
\setlength{\floatsep}{5pt}
\setlength{\abovedisplayskip}{1pt}
\setlength{\belowdisplayskip}{1pt}
\setlength{\abovecaptionskip}{1pt}
\setlength{\belowcaptionskip}{1pt}
\setlength{\tabcolsep}{1pt}
\setlength{\dblfloatsep}{5pt}
\setlength{\dbltextfloatsep}{7pt}

\maketitle
\begin{summary}
We propose a novel hue-correction scheme for multi-exposure image fusion (MEF). 
Various MEF methods have so far been studied to generate higher-quality images.
However, there are few MEF methods considering hue distortion
unlike other fields of image processing, due to a lack of a reference
image that has correct hue.
In the proposed scheme, we generate an HDR image as a reference
for hue correction, from input multi-exposure images.
After that, hue distortion in images fused by an MEF method is removed
by using hue information of the HDR one,
on the basis of the constant-hue plane in the RGB color space.
In simulations, the proposed scheme is demonstrated to be effective
to correct hue-distortion caused by conventional MEF methods.
Experimental results also show that
the proposed scheme can generate high-quality images,
regardless of exposure conditions of input multi-exposure images.
\end{summary}
\begin{keywords}
Multi-exposure image fusion, Color correction, HDR image,
Maximally saturated color, Constant-hue plane
\end{keywords}

\section{Introduction}

The low dynamic range (LDR) of the imaging sensors used in modern digital cameras is a major factor preventing cameras from capturing images as good as those with human vision.
This is due to the limited dynamic range that imaging sensors have, and a single shutter speed that is utilized when we take photos. 
For this reason, there are various methods that aim to
improve the quality of captured images by using multiple images.
Most of the methods utilize a set of differently exposed images,
called "multi-exposure images,"
and fuse them to produce an image with high quality \cite{Debevec,Mitsunaga,Reinhard,Fattal,Durand,Drago,Shan,Siku,Gu,Mertens,Nejati,Bruce,Ma,Kinoshita,KinoshitaMEF,kinoshita_kiya_2018}. 
These methods can be classified into two main approaches.
One is to tone-map (TM) a high dynamic range (HDR) image generated
from input multi-exposure images. 
The other is to directly fuse the multi-exposure images
by using a multi-exposure fusion (MEF) method.

The advantage of MEF compared with the former approach
is that it can generate high-quality images \cite{Seo2}.
However, since MEF does not consider a non-linear response of
cameras used when we take input multi-exposure images,
the resulting image is affected by the hue distortion in the input ones \cite{Artit}.

In other fields of image processing,
color-correction and color-preserving methods have already been studied \cite{Yamaguchi,ie1,ie2,ie3,ie4,ie5,ie6,ie7,Seo1,Mantiuk}.
Ueda et al. \cite{Yamaguchi} developed a hue-preserving contrast enhancement method based on a constant-hue plane in the RGB color space.
Ueda's idea is extended to tone-mapping operations for HDR images
by Kinoshita et al. \cite{Seo1}.
In contrast,
there are few color-correction methods for MEF
because we cannot use input images as a reference for color-correction
unlike other fields of image processing
such as tone mapping.

To solve this problem,
Artit et al. \cite{Artit} proposed a hue-correction method for MEF.
The method fuses input multi-exposure images and
generates an HDR image from the same inputs.
Because the HDR image is generated by considering a non-linear camera response,
it has more accurate hue information than input images.
By using hue information of the generated HDR one as a reference,
the method corrects hue of the fused image.
However,
effects of the performance of the HDR image generation
on the hue correction have never been discussed.
In addition,
Artit's method often generates low-quality unclear images when
unclear input multi-exposure images are given.

Because of such background, in this paper,
we propose a novel hue-correction scheme for MEF.
Similarly to Artit's method,
the proposed scheme generates an HDR image for hue correction from
input multi-exposure images.
Generating the HDR image enables us to obtain a color reference for hue correction in the proposed scheme.
The hue correction is performed on the basis of
the constant-hue plane in the RGB color space.
In addition, to improve the image quality when unclear input multi-exposure images are given,
we use scene segmentation-based luminance adjustment (SSLA) \cite{Kinoshita}.
SSLA enables us to generate clear multi-exposure images from unclear ones.
As a result, the proposed scheme can generates high-quality fused images,
regardless of exposure conditions of input images.
Furthermore, we discuss how the performance of HDR image generation affects the performance of hue correction.

To evaluate the effectiveness of the proposed scheme,
we perform two simulations.
In a simulation,
we first confirm effects of the performance of
HDR image generation on hue correction.
Experimental results show that
higher accuracy of HDR image generation
provides higher performance of hue-correction.
In the other simulation,
the proposed scheme is compared with a conventional MEF method
and a TM operation from a HDR image in terms of the hue difference \cite{CIEDE2000} and image quality \cite{TMQI}.
From this simulation,
it is confirmed that the proposed scheme can generate images
with lower hue-distortion and higher quality than the other methods.
Moreover,
the results also show that the use of SSLA in the proposed scheme
enables us to produce high-quality images
even if unclear input multi-exposure images are given.

\section{Related Work}

Here, we summarize typical multi-exposure image fusion (MEF) methods and image processing methods considering color distortion.
After that, our aim is explained.

\subsection{Multi-Exposure Image Fusion}

The purpose of MEF is to produce images that are expected to be more informative and perceptually appearing than any of the input ones by directly fusing photos taken with different exposures.
The differently exposed images are called "multi-exposure images".

Various research works on MEF have so far been reported \cite{Mertens,Nejati,Li,Sakai,Ma,Bruce,Shen,Kinoshita,KinoshitaMEF,kinoshita_kiya_2018}.
Many of the fusion methods provide a final fused image as a weighted average of input multi-exposure images.
Mertens et al. proposed a multi-scale fusion scheme in which contrast, color saturation, and well-exposedness measures are used for computing fusion weights \cite{Mertens}.
In the work by Nejati et al., a base-detail decomposition algorithm is applied to each input image, and decomposed base- and detail-layers are then fused individually \cite{Nejati}.
A fusion method based on sparse representation was also proposed in \cite{Li}.
Furthermore, a method that combines a weighted-average-based method and a sparse-representation-based method is presented in \cite{Sakai} and is used to enhance image details.

These conventional MEF methods can produce high-quality clear images, but they often cause resulting images to be hue-distorted. 
In addition, even if a MEF method does not cause the hue distortion, hue distortion problem still remains because input multi-exposure images usually have different color \cite{Artit}.

\subsection{Image Processing Considering Color Distortion}

To prevent color distortion caused by image processing methods, color-correction and color-preserving methods have already been studied.

In the field of image enhancement, hue-preserving methods have been studied to avoid the color distortion caused by enhancement \cite{Yamaguchi,ie1,ie2,ie3,ie4,ie5,ie6}.
In these methods, a resulting image can preserve color information of the corresponding input image.
Ueda et al. \cite{Yamaguchi} developed a hue-preserving contrast enhancement method based on a constant-hue plane in the RGB color space.
The use of the constant-hue plane enables us to avoid the gamut problem by enhancing contrast on the plane.

Kinoshita et al. \cite{Seo1} extended the idea of the constant-hue plane used in Ueda's method \cite{Yamaguchi} to tone-map HDR images.
Mantiuk et al. \cite{Mantiuk} also proposed a color correction formula for tone mapping based on a relationship between the contrast-compression ratio and the saturation of the image.
In these methods, hue of tone-mapped LDR images is corrected by using hue information of input HDR images as references.

In contrast to image enhancement and tone mapping, there are no reference images for MEF because input multi-exposure images have different colors from each other.
Therefore, there are few color-correction and color-preserving methods \cite{Artit}.

\subsection{Scenario}

As mentioned above, existing MEF methods have two problems in terms of color:
\begin{itemize}
  \item MEF methods may cause color distortion.
  \item  There are no reference images for hue-correction or hue-preserving methods for MEF.
\end{itemize}
The second problem is caused for the reason that input multi-exposure images have different colors from each other.

To solve these problems, Artit et al. proposed a hue-correction method for MEF.
The method first generates an HDR image from input multi-exposure images, by calibrating a camera response function (CRF).
After that, it corrects the hue of an image fused by MEF methods by using hue information of the HDR one as a reference.
However, this method generates a low-quality image when unclear input multi-exposure images are given.
Also, they have never discussed how the accuracy of CRF calibration affects the performance of hue correction.

For these reasons, in this paper, we propose a novel hue-correction scheme for MEF.
The proposed scheme uses scene segmentation-based luminance adjustment (SSLA) \cite{Kinoshita} to improve the image quality when unclear input multi-exposure images are given.
In addition, we discuss the effects of the accuracy of CRF calibration on the performance of hue correction.

\begin{figure*}[t]
  \centering
\includegraphics[keepaspectratio, scale=0.43]{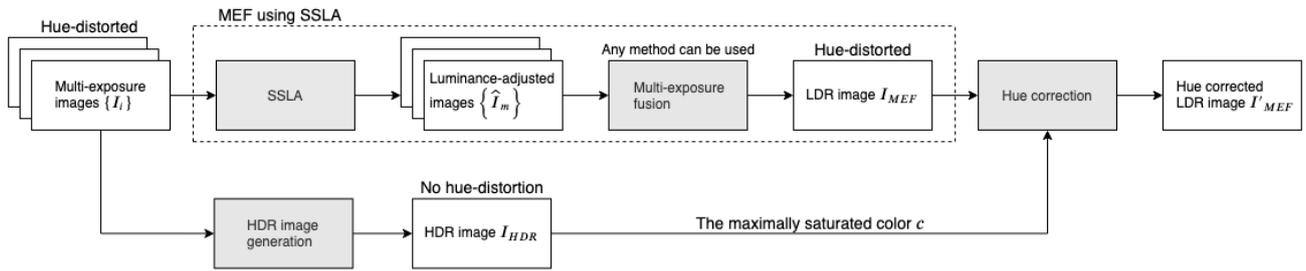}
\caption{Overview of proposed scheme\label{fig:overview}}
\end{figure*}

\section{Technical Background}

The expression of hue considered in this paper is the hue as defined in HSI color space \cite{ie2}, which corresponds to the maximally saturated color on the constant-hue plane.
In this section, the constant-hue plane and hue distortion in MEF is first explained, then MEF using SSLA and HDR image generation are summarized. In the proposed scheme, MEF using SSLA and an HDR image generation method are used for multi-exposure fusion and hue-correction, respectively.

\subsection{Constant-Hue Plane in the RGB Color Space}

\begin{figure}[t]
  \centering
\includegraphics[keepaspectratio, scale=0.16]{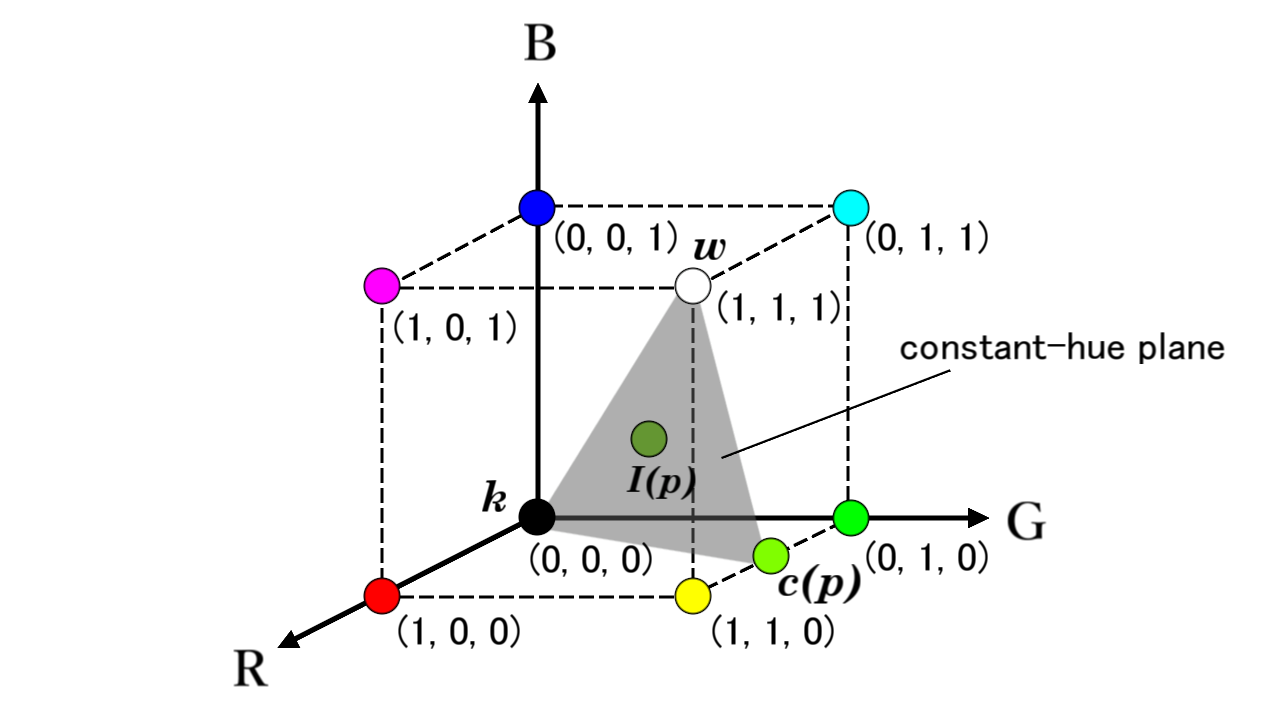}
\caption{Conceptual diagram of RGB color space. Gray triangle denote constant hue plane for pixel value $\bm{I} (p)$\label{fig:rgb_color_space}}
\end{figure}

Each pixel of RGB color image $\bm{I}$ can be represented as $\bm{I} (p) \in [0,1]^3$, where the R, G, and B components of pixel $\bm{I}(p)$ are written as $I_r(p), I_g(p)$ and $I_b(p)$, respectively.
In the RGB color space, a set of pixels which has the same hue forms a plane, called "constant-hue plane," as shown in Fig.\ref{fig:rgb_color_space}.
The shape of each constant-hue plane is a triangle whose vertices are white $\bm{w}=(1,1,1)$, black $\bm{k}=(0,0,0)$, and maximally saturated color $\bm{c}(p)$.
The maximally saturated color $\bm{c} (p) =(c_r (p) , c_g (p) , c_b (p) )$, which has the same hue as that of $\bm{I}(p)$, is calculated by
\[
	c_r (p)  = \frac{I_r (p) -\min{(\bm{I} (p) )}}{\max{(\bm{I} (p) )}-\min{(\bm{I} (p) )}},
\]
\begin{equation}
\label{eq:c}
	c_g (p)  = \frac{I_g (p) -\min{(\bm{I} (p) )}}{\max{(\bm{I} (p) )}-\min{(\bm{I} (p) )}},
\end{equation}
\[
	c_b (p)  = \frac{I_b (p) -\min{(\bm{I} (p) )}}{\max{(\bm{I} (p) )}-\min{(\bm{I} (p) )}},
\]
where $\max(\cdot)$ and $\min(\cdot)$ are functions that return the maximum and minimum elements of pixel $\bm{I} (p) $, respectively.

On the constant hue plane, pixel $\bm{I}(p)$ can be represented as a linear combination as
\begin{equation}
\label{eq:x}
	\bm{I} (p) =a_w (p)  \bm{w}+a_k (p)  \bm{k}+a_c (p)  \bm{c} (p) ,
\end{equation}
where 
\[
	a_w (p)  = \min(\bm{I} (p) ),
\]
\begin{equation}
\label{eq:a}
	a_c (p)  = \max(\bm{I} (p) )-\min(\bm{I} (p) ),
\end{equation}
\[
	a_k (p)  = 1-\max(\bm{I} (p) ).
\]
Since $\bm{w},\bm{k},\bm{c}(p)$ and $\bm{I}(p)$ exist on the plane and $\bm{I}(p)$ is an interior point of $\bm{w},\bm{k}$ and $\bm{c}(p)$, the following equations hold:
\begin{equation}
\label{eq:a_condition1}
	a_w(p) +a_k(p) +a_c(p) =1,
\end{equation}
\begin{equation}
\label{eq:a_condition2}
	0 \leq a_w(p) ,a_k(p) ,a_c(p) \leq 1.
\end{equation}

\subsection{Hue Distortion in Multi-Exposure Images}

\begin{figure}[t]
  \centering
\includegraphics[keepaspectratio, scale=0.26]{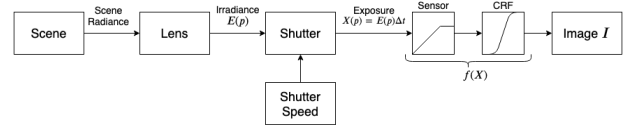}
\caption{Imaging pipeline of digital camera\label{fig:camera}}
\end{figure}

Hue distortion in input multi-exposure images is caused by a non-linear response of a digital camera that are used for capturing the images.
Figure \ref{fig:camera} shows a typical imaging pipeline for a digital camera \cite{camera}.
The radiant power density at the sensor, called irradiance $\bm{E}(p)$, is integrated over the time $\Delta t$ the shutter is open, producing an energy density, commonly referred to as exposure $\bm{X}(p) = (X_r(p), X_g(p), X_b(p))$.
$X_r(p), X_g(p)$ and $X_b(p)$ are equivalent to R, G and B components of exposure values, respectively.
If the scene is static during this integration, exposure $\bm{X}(p)$ can be written simply as the product of irradiance $\bm{E}(p)$ and integration time $\Delta t$ (referred to as “shutter speed”):
\begin{equation}
\label{eq:exposure}
\bm{X}(p) = \bm{E}(p) \Delta t.
\end{equation}
A pixel value $\bm{I}(p)$ in captured image $I$ is given by
\begin{equation}
\label{eq:crf_rgb}
\bm{I}(p) = f(\bm{X}(p)) = \left(
\begin{array} {c}
    f_r(X_r(p) ) \\
    f_g(X_g(p) ) \\
    f_b(X_b(p) )  \\
  \end{array}
  \right), 
\end{equation}
where $f_r, f_g,$ and $f_b$ are non-linear functions combing sensor saturation and camera response functions (CRF) for red, green, and blue components.
The CRF represents the processing in each camera which makes the final image $I$ look better.

From Eq.(\ref{eq:c}), maximally saturated color $\bm{c}(p)$ (i.e., hue) for $\bm{I}(p)$ depends only on the ratio of $I_r(p) , I_g(p) $, and $I_b(p) $.
R, G and B components are independently converted by the non-linear functions. 
Thus, the ratio will be changed from that of $\bm{E}$.
By the influence, the image $\bm{I}$ has hue distortion.


\subsection{Hue Distortion caused by MEF}

As in the Section 3.2, hue distortion occurs when the ratio of R, G, and B components is changed.
For this reason, even if input images do not have hue distortion,
MEF methods that change the ratio cause fused images to be hue-distorted.

\subsection{Multi-Exposure Fusion Using SSLA}

SSLA \cite{Kinoshita} enables us to generate clear multi-exposure images from unclear multi-exposure images.
It can improve the quality of resulting fused images.
The procedure of SSLA is shown as follows:

\begin{enumerate}[label=(\alph*)]
\item Enhance local contrast of input multi-exposure images $\{\bm{I}_i\}$ by
  \begin{equation}
    \label{eq:enhancement}
    L'_i(p) = \frac{L^{2}_i(p)}{L_{ai}(p)},
  \end{equation}
  where $L_i(p)$ is the luminance value of the $i$-th input image $\bm{I}_i$
  at pixel $p$, and $L_{ai}(p)$ is the local average of luminance $L_i$
  around pixel $p$.

\item Separate a scene in multi-exposure images
 into $M$ areas $\mathrm{P}_1, \mathrm{P}_2, \cdots, \mathrm{P}_M$,
 where each of them has a specific brightness range of the scene.
 For the scene segmentation,
 a Gaussian mixture distribution is utilized to model the luminance distribution of all input images.
 After that, pixels are classified into $M$ areas
 $\mathrm{P}_1, \mathrm{P}_2, \cdots, \mathrm{P}_M$
 by using a clustering algorithm based on
 a Gaussian mixture model (GMM) \cite{GMM}.

\item Obtain scaled luminance $L''_m$ by
  \begin{equation}
    \label{eq:L_m}
    L''_m(p) = \alpha _m L'_n(p),
  \end{equation}
  where parameter $\alpha _m > 0$ indicates
  the degree of adjustment for the $m$-th scaled luminance $L''_m$.
  The degree of adjustment is calculated
  so that $L''_m$ clearly represents area $\mathrm{P}_m$, as
  \begin{equation}
    \label{eq:alpha_m}
    \alpha _m = \frac{0.18}{g(L'_n | \mathrm{P}_m)},
  \end{equation}
  where $g(L'_n | \mathrm{P}_m)$ is
  the geometric mean of luminance $L'_n$ on area $\mathrm{P}_m$.
  Since a smaller value for parameter $\alpha _m$ is better,
  $n$ is chosen as
  \begin{equation}
    \label{eq:n}
    n = \phi(m) = \argmin_{j} (0.18 - g(L'_n | \mathrm{P}_m))^2.
  \end{equation}

\item Tone map scaled luminance value $L''_m$ by
  \begin{align}
    \label{eq:tm}
    \hat{L}_m(p) &= f_m (L''_m(p)),\\
    f_m(t) &= \frac{t}{1 + t}\left(1 + \frac{t}{l^2_m} \right),
  \end{align}
  where we set parameter $l_m$ as
  $l_m = \max L''_m$ as in \cite{Kinoshita}.

\item Combine a set ${\hat{L}_m}$ of luminance adjusted by the SSLA
  with input multi-exposure images ${I_n}$
  to obtain adjusted images ${\hat{I}_m}$ as follows
  \begin{equation}
    \label{eq:adjusted_image}
    \hat{\bm{I}}_m(p) = \frac{\hat{L}_m(p)}{L_{\phi (m)}(p)} \bm{I}_{\phi (m)}(p),
  \end{equation}
  where eq.(\ref{eq:n}) is utilized
  to associate each ${\hat{L}_m}$ with an input image $\bm{I}_n$.
\end{enumerate}

Generated multi-exposure images $\{\hat{\bm{I}}_m\}$ can be fused into $\bm{I}_{\rm{MEF}}$ by using an MEF method $\mathcal{F}$, as
\begin{equation}
\label{eq:mef}
\bm{I}_{\rm MEF} = \mathcal{F}(\hat{\bm{I}}_1, \hat{\bm{I}}_2, \cdots, \hat{\bm{I}}_m).
\end{equation}
Here, we can use any existing MEF methods $\mathcal{F}$.
Mertens’ MEF method \cite{Mertens} is used in this paper.

\subsection{HDR Image Generation}

HDR image generation \cite{Debevec, Mitsunaga} methods can calculate irradiance $\bm{E}$ by removing the non-linearity of $f$.
For this reason, we utilize an HDR image generation method for removing hue distortion in multi exposure images and distortion caused by MEF.

In accordance with Eq.(\ref{eq:exposure}) and Eq.(\ref{eq:crf_rgb}), the pixel value $\bm{I}_i (p)$ in captured image $\bm{I}_i$ is calculated as
\begin{equation}
\label{eq:x_captured}
	\bm{I}_i (p) = f(\bm{E}(p) \Delta t_i),
\end{equation}
where $\Delta t_i$ is an integration time for the $i$-th image $\bm{I}_i$.
From Eq.(\ref{eq:x_captured}),  the irradiance map is calculated by
\begin{equation}
\label{eq:loge}
	\ln \bm{E}_i (p) = \ln f^{-1} (\bm{I}_i (p)) - \ln \Delta t_i,
\end{equation}
where $f^{-1}$ is an inverse function of $f$.
An HDR image $\hat{\bm{I}}_{\rm{HDR}}$ is obtained by
\begin{equation}
\label{eq:hdr_pixel}
	\hat{\bm{I}}_{\rm{HDR}}(p) = \exp \left( \frac{\sum_{i=1}^{N} \omega(\bm{I}_i(p)) \ln \bm{E}_i}{\sum_{i=1}^{N} \omega(\bm{I}_i(p))} \right),
\end{equation}
where $\omega (\cdot)$ is a weighting function.

Since $f^{-1}$ is generally unknown, $f^{-1}$ is estimated from input multi-exposure images.
Typical methods for estimating $f^{-1}$ are Debevec's method \cite{Debevec} and Mitsunaga's method \cite{Mitsunaga}.
The estimation accuracy (i.e., the performance of removing the non-linearity of $f$) directly affects the performance of hue correction.
The effect of estimation accuracy on hue correction will be discussed in Session 5.

\section{Proposed Hue-Correction Scheme}

Figure \ref{fig:overview} shows an overview of our hue-correction scheme for multi-exposure fusion.
The proposed scheme consists of MEF using SSLA, HDR image generation, and hue correction.
Because the color of input multi-exposure images is distorted by non-linear CRF (See 3.2), 
we calculate a color reference for hue correction by generating an HDR image.
In the proposed scheme, we can use any existing MEF methods.

In this section, hue correction is first summarized, and the proposed procedure is then explained.

\begin{figure*}[h]
  \centering
  \subfigure[-4{[}EV{]}]{
    \includegraphics[scale=0.15]{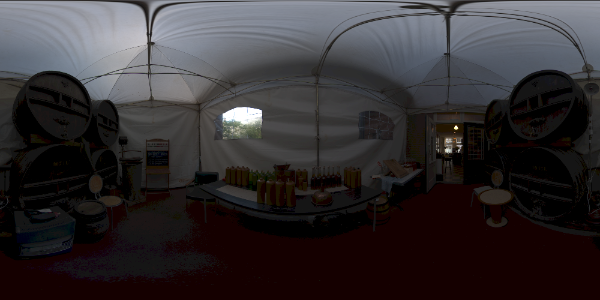}}
    \subfigure[-2{[}EV{]}]{
    \includegraphics[scale=0.15]{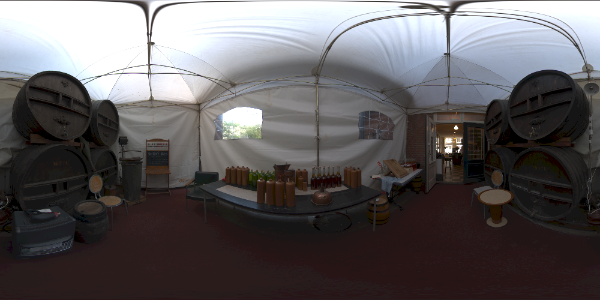}}
    \subfigure[0{[}EV{]}]{
    \includegraphics[scale=0.15]{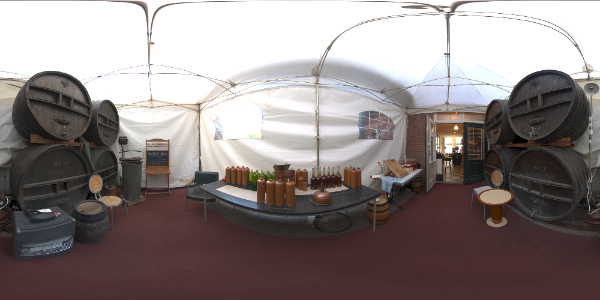}}
    \subfigure[2{[}EV{]}]{
    \includegraphics[scale=0.15]{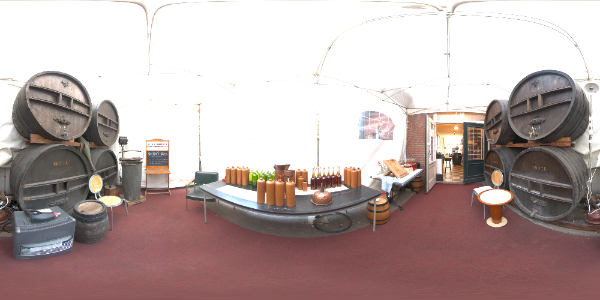}}
    \subfigure[4{[}EV{]}]{
    \includegraphics[scale=0.15]{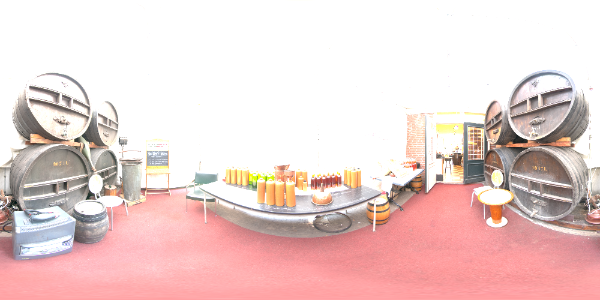}}
  \caption{Example of input multi-exposure images $\{\bm{I}_i\}$ \label{fig:input_images}}
\end{figure*}

\subsection{Hue Correction}

In our hue correction, we aim to remove hue distortion included in fused image $\bm{I}_{\rm{MEF}}$ by using hue-information of $\bm{I}_{\rm{HDR}}$.
In accordance with Eq.(\ref{eq:x}), a pixel value $\bm{I}_{\rm{MEF}}(p)$ of fused image $\bm{I}_{\rm{MEF}}$ is represented as
\begin{equation}
\label{eq:x_mef}
	\bm{I}_{\rm{MEF}} (p) = a_w(p) \bm {w}+ a_k(p) \bm{k}+ a_c(p) \bm{c}(p).
\end{equation}
Likewise, a pixel value $\bm{I}_{\rm{HDR}}(p)$ of HDR image $\bm{I}_{\rm{HDR}}$ is represented as
\begin{equation}
\label{eq:x_hdr}
\begin{split}
	\bm{I}_{\rm{HDR}} (p) = a_{w,\rm{HDR}}(p) \bm {w}+ a_{k,\rm{HDR}}(p) \bm{k} \\
	+ a_{c,\rm{HDR}}(p) \bm{c}_{\rm{HDR}}(p),
\end{split}
\end{equation}
where $a_{w,\rm{HDR}}(p), a_{k,\rm{HDR}}(p), a_{c,\rm{HDR}}(p),$ and $ \bm{c}_{\rm{HDR}}(p)$ are calculated from HDR image $\bm{I}_{\rm{HDR}}$ by using Eq.(\ref{eq:c}) and Eq.(\ref{eq:a}).
Note that $a_{w,\rm{HDR}}(p), a_{k,\rm{HDR}}(p),$ and $a_{c,\rm{HDR}}(p)$ do not satisfy Eq.(\ref{eq:a_condition2}).
Substituting $\bm{c}_{\rm{HDR}}(p)$ for $\bm{c}(p)$, a hue-corrected pixel value $\bm{I}'_{\rm{MEF}}(p)$ is calculated as follows
\begin{equation}
\label{eq:x'}
	\bm{I}'_{\rm{MEF}}(p) = a_w(p) \bm {w}+ a_k(p) \bm{k}+ a_c(p) \bm{c}_{\rm{HDR}}(p).
\end{equation}
$\bm{I}'_{\rm{MEF}}(p)$ and $\bm{I}_{\rm{HDR}}(p)$ are on the same constant-hue plane because they have the same maximally saturated color $\bm{c}_{\rm{HDR}}$.

The constant-hue plane in the RGB color space is based on the HSI color space \cite{Yamaguchi}.
The hue correction that directly replaces the hue value in the HSI color space does not consider the color gamut of the RGB color space.
For this reason, resulting pixel values may be out of the color gamut of the RGB color space.
In contrast, the hue correction in the proposed scheme ensures that the resulting pixel values are in the RGB color gamut
because the hue correction based on the constant-hue plane in the RGB color space considers it.

\subsection{Proposed Procedure}

By using the proposed scheme, MEF is performed as follows (See also Fig.\ref{fig:overview}):

\begin{enumerate}
\item Generate a fused image $\bm{I}_{\rm{MEF}}$ from input multi-exposure images $\{ \bm{I}_i \}$ by MEF using SSLA in accordance with Eq.(\ref{eq:enhancement}) to Eq.(\ref{eq:mef}).
\item Generate an HDR image $\hat{\bm{I}}_{\rm{HDR}}$ from $\{ \bm{I}_i \}$ by using Eq.(\ref{eq:loge}) and Eq.(\ref{eq:hdr_pixel}).
\item Calculate coefficients $a_w(p), a_k(p)$, and $a_c(p)$ for each pixel value in fused image $\bm{I}_{\rm{MEF}}$ in accordance with Eq.(\ref{eq:a}).
\item Calculate the maximally saturated color $ \bm{c}_{\rm{HDR}} (p)$ for each pixel in HDR image $\hat{\bm{I}}_{\rm{HDR}}$ by using Eq.(\ref{eq:c}).
\item Obtain hue-corrected image $\bm{I}'_{\rm{MEF}}$ in accordance with Eq.(\ref{eq:x'}).
\end{enumerate}

\section{Simulation}

We performed two simulations to confirm the effectiveness of the proposed scheme.

\subsection{Objective Metrics}

We evaluated images fused with/without the proposed scheme in terms of hue distortion and their quality.
For evaluating hue distortion in fused images, we used the hue difference $\Delta H$ in CIEDE2000\cite{CIEDE2000}.
The hue difference $\Delta H$ between a fused image and a reference image was first calculated for each pixel, and then the average of $\Delta H$ was used as a hue-difference score.
For evaluating the quality of fused images, we utilized Tone Mapped image Quality Index (TMQI)\cite{TMQI}, which measures structural fidelity and statistical naturalness of a tone-mapped image from an HDR one.
A higher TMQI score indicates higher quality.

These metrics need a reference image for evaluating a fused image.
However, we cannot prepare reference images when multi-exposure images taken with cameras are used as inputs.
Therefore, in this paper, we generated input multi-exposure images from HDR images.

\subsection{Dataset}

In simulations, a set of input multi-exposure images $\{\bm{I}_i\}$ was generated from a set of HDR ones.
These HDR images were used as reference images for calculating both $\Delta H$ and TMQI.
The following is the procedure for generating multi-exposure images $\{\bm{I}_i\}$.
\begin{enumerate}
    \item Generate an exposure $\bm{X}_i$ with exposure value $v_i$ [EV] from an HDR image $\bm{I}_{\rm HDR}$ as
\begin{equation}
    \bm{X}_i(p) = 2^{v_i}\frac{0.18}{g(L(p)|\rm{P})} \bm{I}_{\rm HDR}(p),
\end{equation}
where $\rm{P}$ indicates the whole area on the image.
    \item Clip the exposure $\bm{X}_i(p)$ into a range from 0 to 1.
    \item Apply a gamma curve ($\gamma = 2.2$) as a non-linear function $f$ to $\bm{X}_i$ as
  \begin{equation}
\label{eq:dataset}
\bm{I}_{i,f}(p) = \left(
\begin{array} {c}
    f_r(X_{i,r}(p) ) \\
    f_g(X_{i,g}(p) ) \\
    f_b(X_{i,b}(p) )  \\
  \end{array}
  \right) =  \left(
\begin{array} {c}
    X_{i,r}(p)^{1/2.2} \\
    X_{i,g}(p)^{1/2.2} \\
    X_{i,b}(p)^{1/2.2}  \\
  \end{array}
  \right).
\end{equation}
    \item Obtain 8-bit pixel values $\bm{I}_i(p)$ as
    \begin{equation}
    \bm{I}_i(p) = {\rm round}(\bm{I}_{i,f}(p) \cdot 255),
    \end{equation}
    where round$(\cdot)$ rounds each element of a vector to nearest integer value.
\end{enumerate}{}

We used 140 HDR images which were selected from a database\cite{HDR}.
In Step.2, $\{v_i\}= \{-4,-2,0,2,4\}$ [EV] were used.
Figure \ref{fig:input_images} shows an example of the generated input images $\{\bm{I}_i\}$.

\subsection{Simulation 1: Accuracy of Estimating a Non-Linear Function $f$ and Effects on Hue-Correction}

To confirm the effect of the estimation accuracy of a non-linear function $f$ on hue-correction, we applied two HDR generation methods, Mitsunaga's method \cite{Mitsunaga} and Debevec's method \cite{Debevec}, to proposed scheme.

\begin{figure}[t]
    \centering
    	\subfigure[Mitsunaga's method]{
		\label{subfig:debevec}
		\includegraphics[width = 0.48\columnwidth]{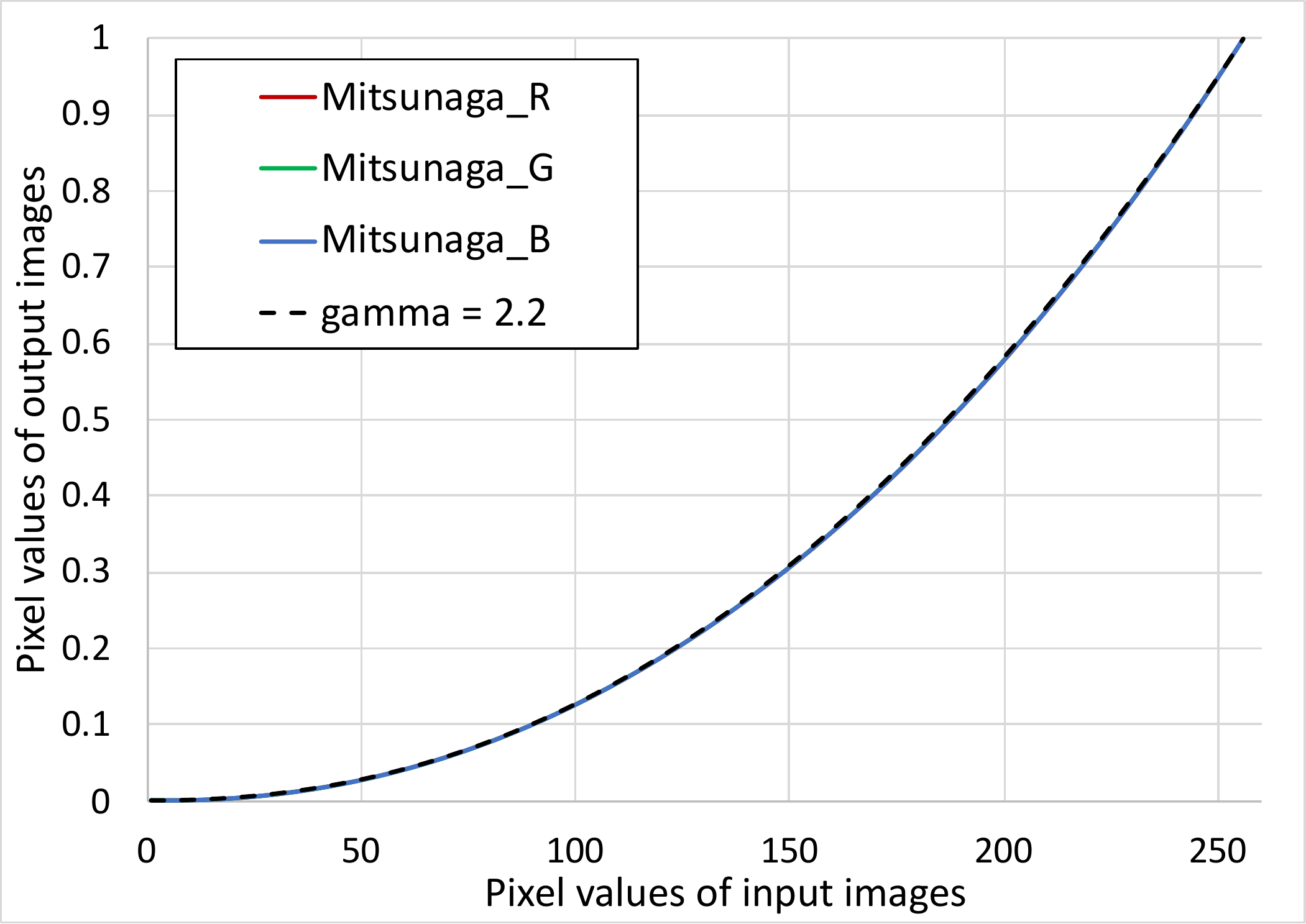}}
	\subfigure[Debevec's method]{
		\label{subfig:mitsunaga}
		\includegraphics[width = 0.48\columnwidth]{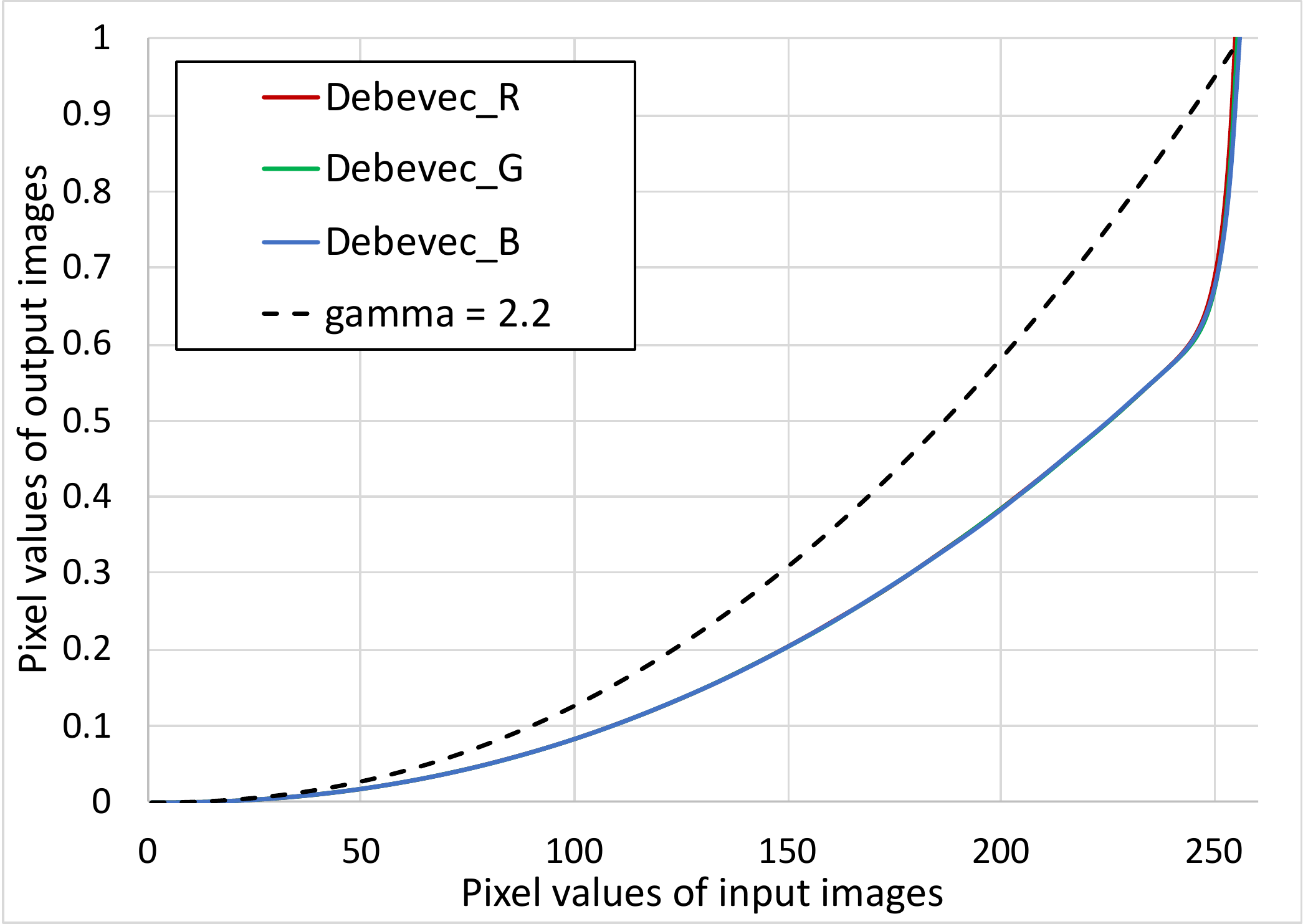}}
    \caption{Estimation of inverse non-linear function $f^{-1}$ }
    \label{fig:estimated_crf}
\end{figure}

Figure \ref{fig:estimated_crf} shows gamma curve $f$ (ground truth) and its estimations.
From Figure \ref{fig:estimated_crf}, the Mitsunaga's method provided a better estimation than Debevec's method.
Table \ref{tab:crf_mse} shows MSE scores between the gamma curve $f$ and its estimations, which were averaged over all 140 image sets.
From Table \ref{tab:crf_mse}, the accuracy of Mitsunaga's method was higher than Debevec's method similarly to Fig.\ref{fig:estimated_crf}.

\begin{figure}[t]
    \centering
    	\subfigure[Box-plot of hue difference]{
		 \includegraphics[width = 0.48\columnwidth]{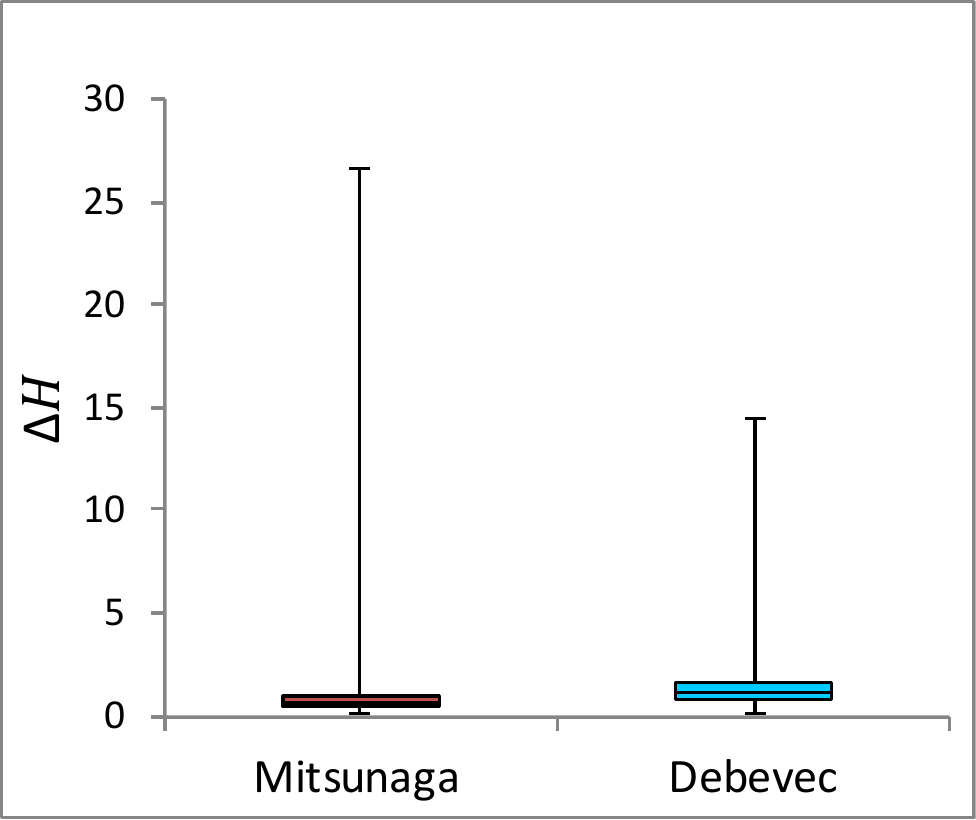}}
	\subfigure[Zoom-in of (a) without whiskers]{
		 \includegraphics[width = 0.48\columnwidth]{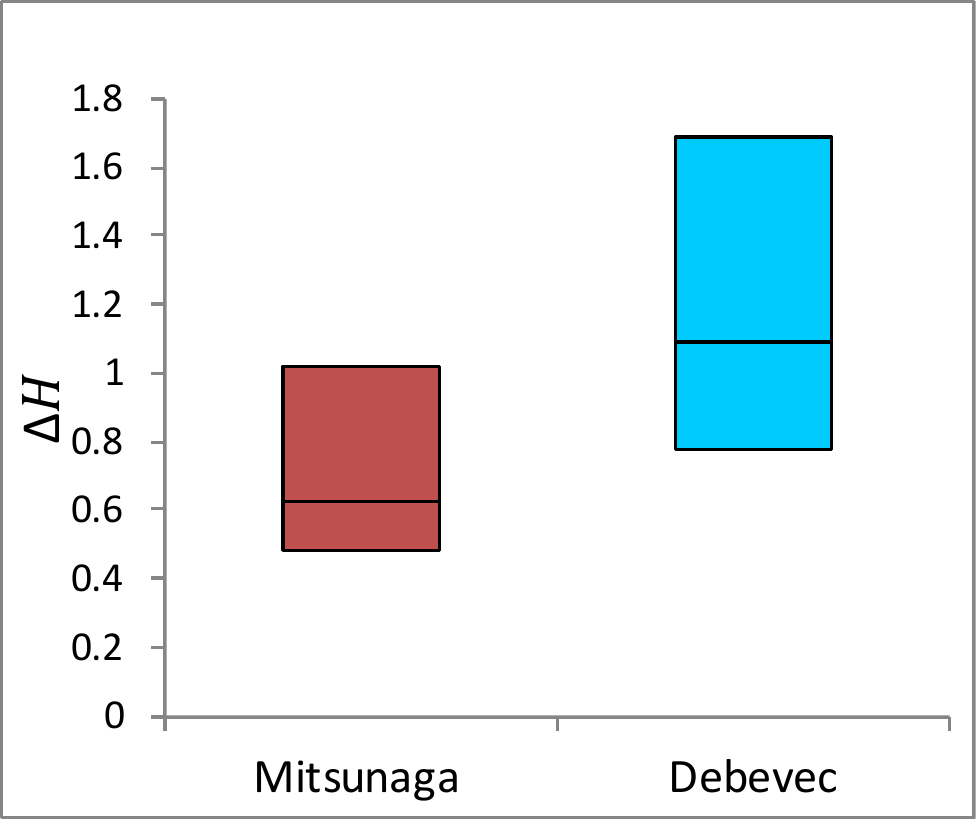}}
\caption{Effect of estimation accuracy of non-linear function $f$ on hue correction. Boxes span from the first to the third quartile referred to as $Q_1$ and $Q_3$, and whiskers show maximum and minimum values in the range of $[Q_1-1.5(Q_3-Q_1),Q_3+1.5(Q_3-Q_1)]$. Band inside box indicates median. \label{fig:crf_tm}}
\end{figure}

Figure \ref{fig:crf_tm} shows box-plots of hue-difference scores for resulting images of the proposed scheme under the use of the two HDR generation methods.
From Figure \ref{fig:crf_tm}, in the case using  Mitsunaga's method,
the average hue differences $\Delta H$ are lower than those of Debevec's method for most images,
while the maximum is higher than that of Debevec's method.
In such a case that Mitsunaga's method does not work well,
the input multi-exposure images contained many saturated white pixels.
Therefore, these results show that Mitsunaga's method is capable for reproducing colors for many input images,
although Debevec's method is capable for input images containing many saturated white pixels.

For these reasons, Mitsunaga's method, which works well for many input images, was used for HDR image generation in the proposed scheme.

\begin{table}[t]
\centering
\caption{MSE scores between non-linear function $f$ and its estimation. MSE score were averaged over all 140 input image set.}
\label{tab:crf_mse}
\begin{tabular}{cccc}
\hline
 & R & G & B \\ \hline
Mitsunaga & 0.00000905 & 0.00000848 & 0.00000962 \\
Debevec & 0.03131296 & 0.03069316 & 0.03054029 \\ \hline
\end{tabular}
\end{table}

\subsection{Simulation 2: Comparison with Existing MEF Methods}

In this simulation, we compared the proposed scheme with existing MEF methods and tone-mapping methods under various input-conditions.
The compared methods were:
\begin{enumerate}[label=(\Alph*)]
  \item Fusing by Mertens' MEF method \cite{Mertens} 
  \item Tone-mapping by Fattal's TM operator \cite{Fattal} from a HDR image $\hat{\bm{I}}_{\rm{HDR}}$ generated from input images  $\{ \bm{I}_i \}$
  \item Fusing by Mertens' MEF using SSLA \cite{Kinoshita}
  \item Fusing by the proposed scheme with Mertens' MEF method
\end{enumerate}

\begin{figure}[t]
    \centering
    	\subfigure[$\{v_i\}=\{-4,-2,0,2,4\}${[}EV{]}]{
		\label{subfig:dh_case4}
		\includegraphics[width = 0.48\columnwidth]{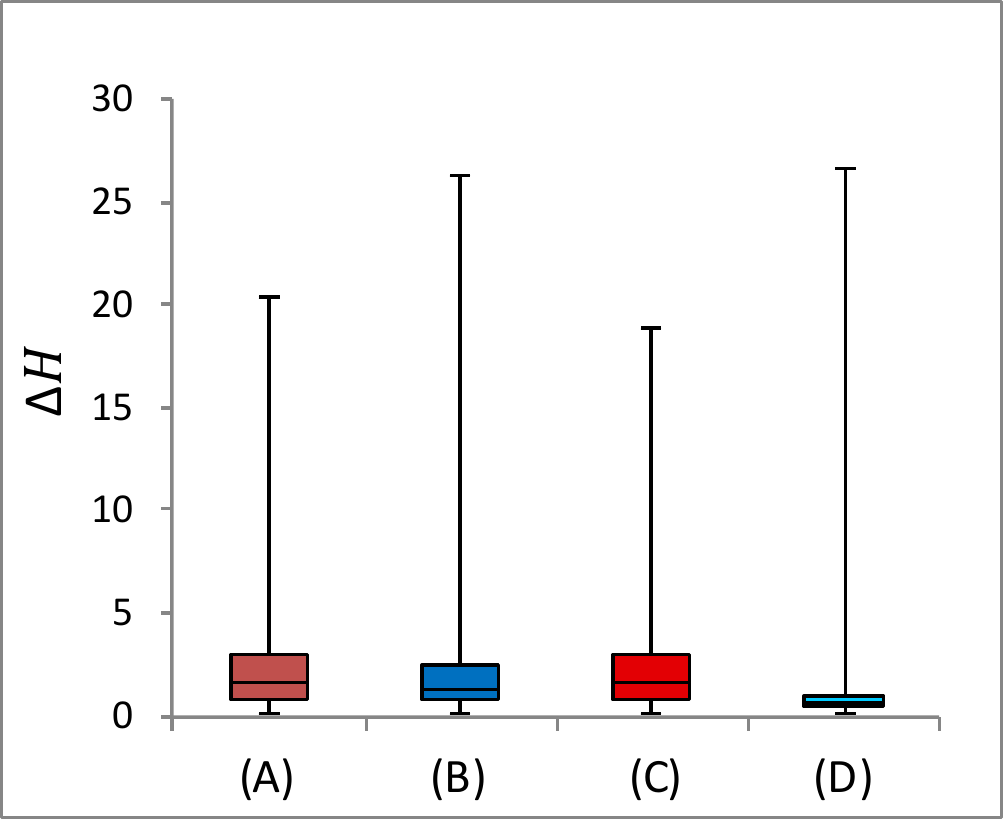}} \\
	\subfigure[$\{v_i\}=\{-4,-2,0\}${[}EV{]}]{
		\label{subfig:dh_case5}
		\includegraphics[width = 0.48\columnwidth]{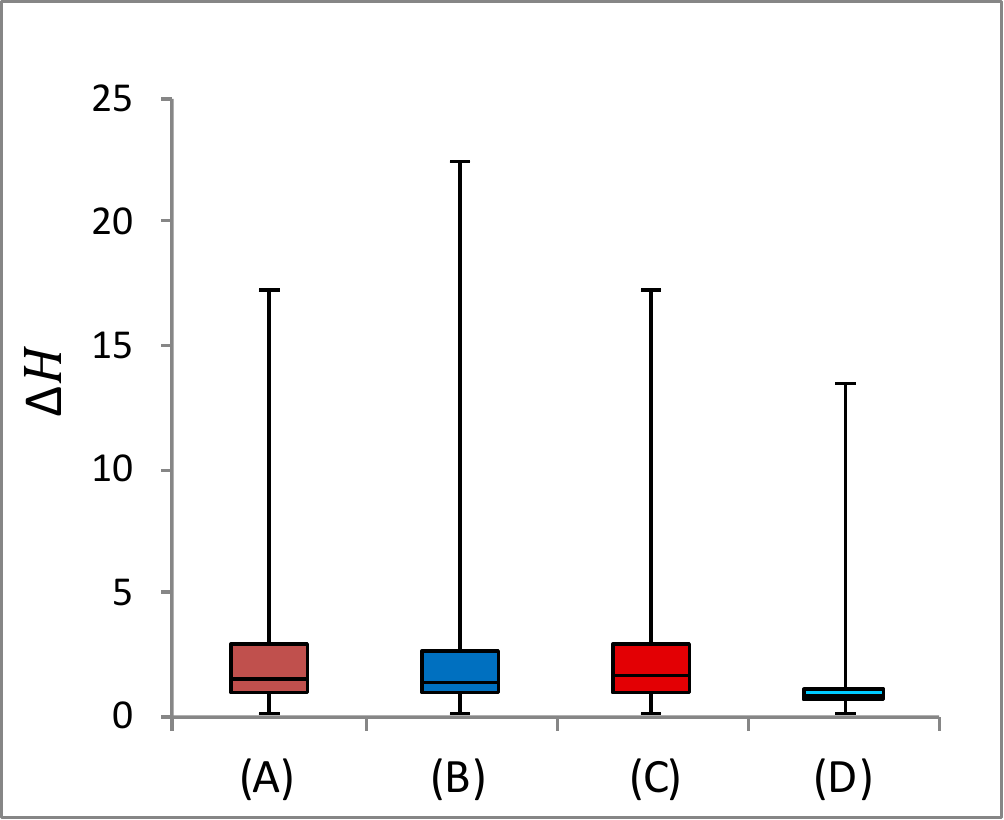}}
	\subfigure[$\{v_i\}=\{0,2,4\}${[}EV{]}]{
		\label{subfig:dh_case6}
		\includegraphics[width = 0.48\columnwidth]{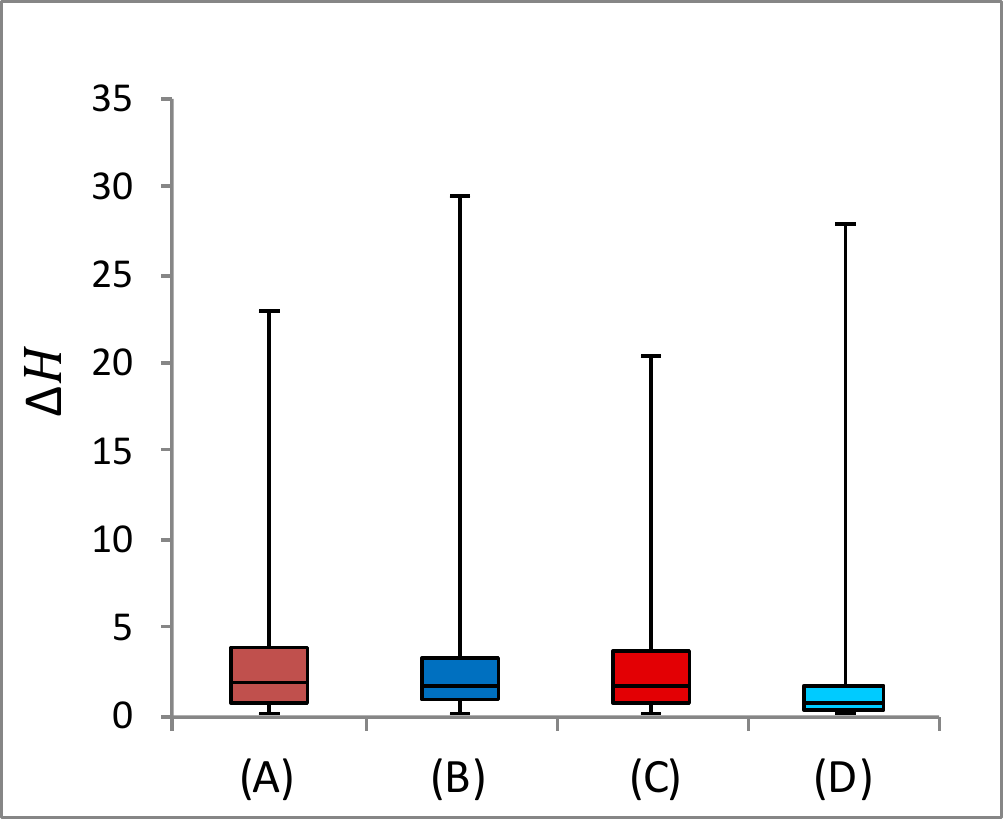}}
    \caption{Experimental result ($\Delta H$). (A) Mertens' MEF \cite{Mertens}, (B) Tone-mapped from HDR image by Fattal's method \cite{Fattal}, (C) Only SSLA, (D) Proposed. Boxes span from the first to the third quartile referred to as $Q_1$ and $Q_3$, and whiskers show maximum and minimum values in the range of $[Q_1-1.5(Q_3-Q_1),Q_3+1.5(Q_3-Q_1)]$. Band inside box indicates median.
    \label{fig:box_dh1}}
\end{figure}

Figure \ref{fig:box_dh1} shows box-plots of hue difference scores for 140 fused images.
The difference among Figs.\ref{subfig:dh_case4}, \ref{subfig:dh_case5}, and \ref{subfig:dh_case6} are exposure values $\{v_i\}$ that input multi-exposure images have.
From Fig.\ref{fig:box_dh1}, focusing on the average hue difference scores,
the proposed scheme outperformed the other three methods in terms of the hue distortion in all three conditions.
Figure \ref{subfig:dh_case4} and \ref{subfig:dh_case6} also show that
the maximum score of the proposed scheme is higher than the other methods
when multi-exposure images having many saturated white pixels are given.
In contrast, Fig.\ref{subfig:dh_case5} illustrates that
scores of the proposed scheme including the maximum are better than the other methods in all conditions
when input images did not contain many saturated white pixels.
Therefore, when the input images contain many saturated white pixels,
the performance of the proposed scheme can be improved by removing some bright input images such as ones having 4[EV] and 2[EV].

\begin{figure}[t]
    \centering
    	\subfigure[$\{v_i\}=\{-4,-2,0,2,4\}${[}EV{]}]{
		\label{subfig:tmqi_case4}
		\includegraphics[width = 0.48\columnwidth]{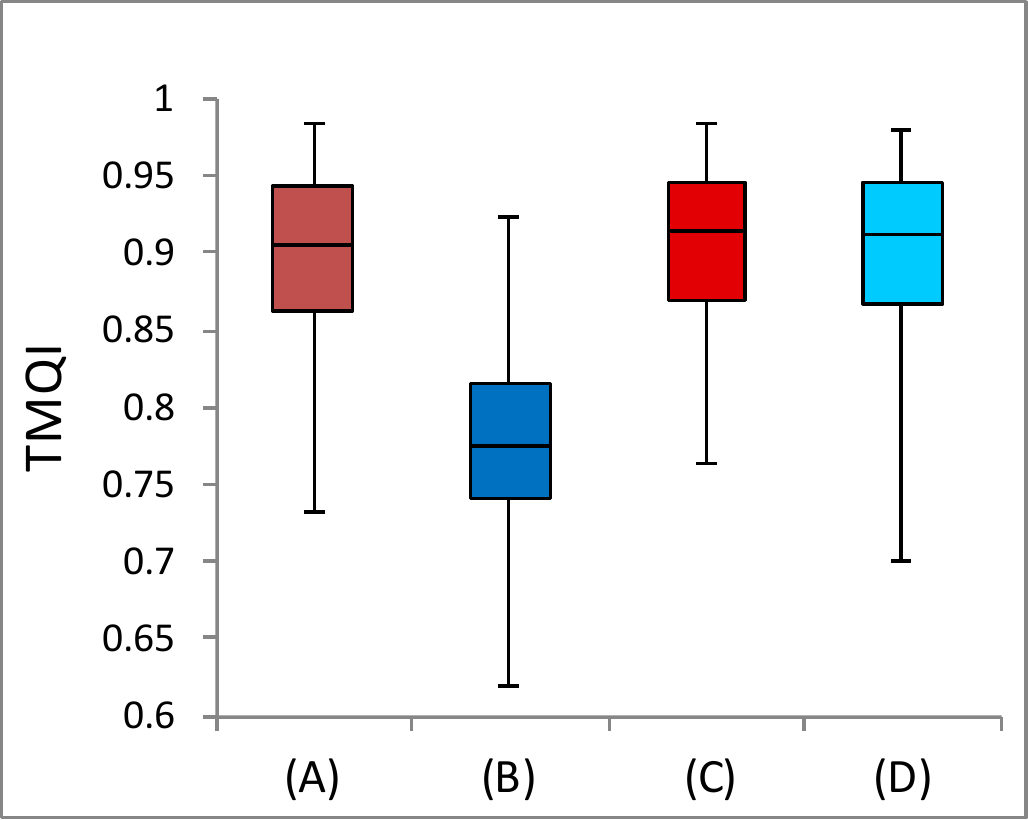}} \\
	\subfigure[$\{v_i\}=\{-4,-2,0\}${[}EV{]}]{
		\label{subfig:tmqi_case5}
		\includegraphics[width = 0.48\columnwidth]{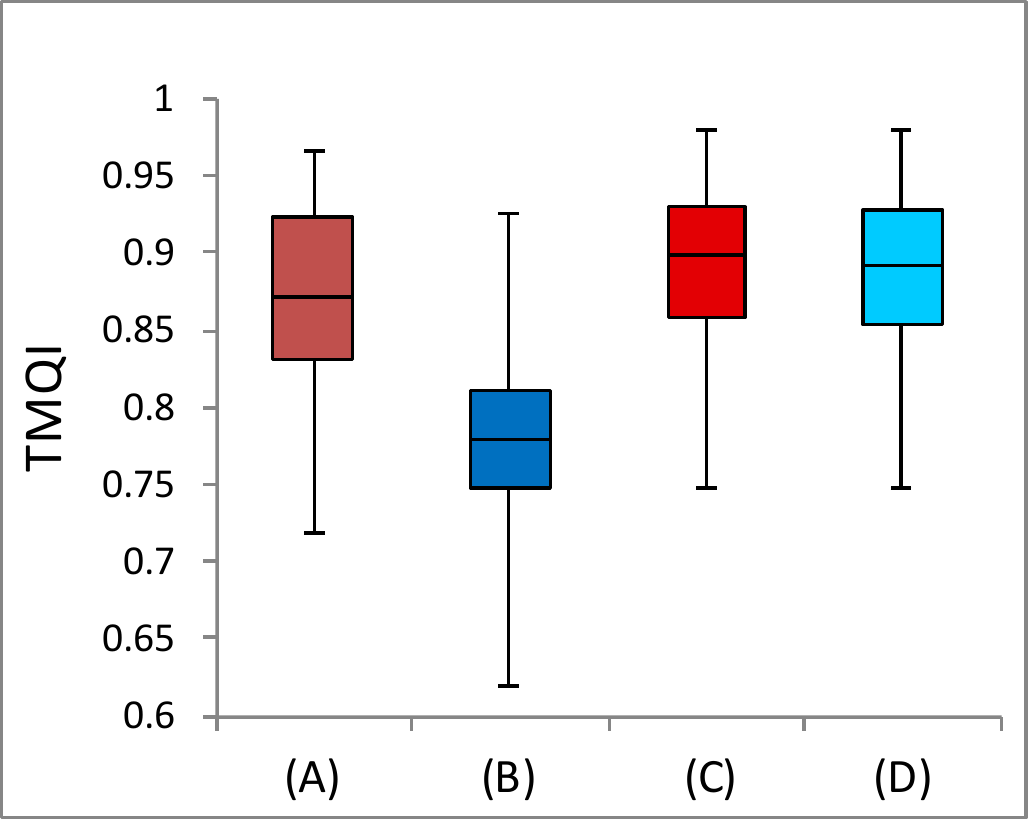}}
	\subfigure[$\{v_i\}=\{0,2,4\}${[}EV{]}]{
		\label{subfig:tmqi_case6}
		\includegraphics[width = 0.48\columnwidth]{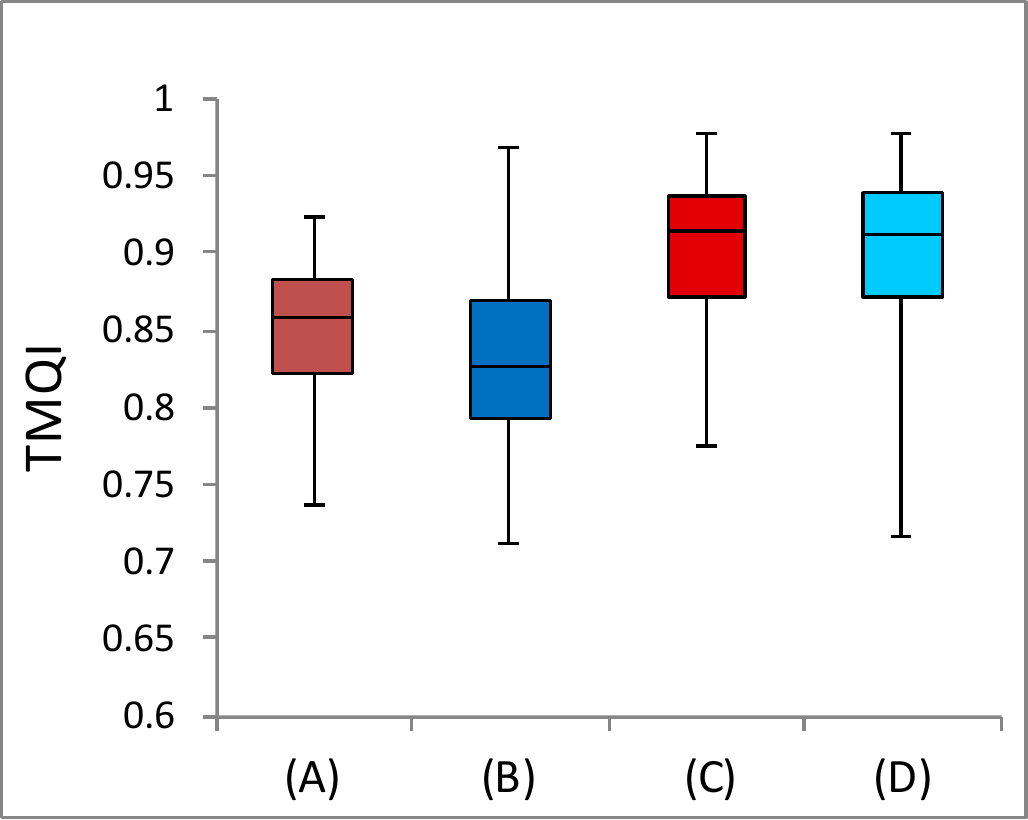}}
    \caption{Experimental result (TMQI). (A) Mertens' MEF \cite{Mertens}, (B) Tone-mapped from HDR image by Fattal's method \cite{Fattal}, (C) Only SSLA, (D) Proposed. Boxes span from the first to the third quartile referred to as $Q_1$ and $Q_3$, and whiskers show maximum and minimum values in the range of $[Q_1-1.5(Q_3-Q_1),Q_3+1.5(Q_3-Q_1)]$. Band inside box indicates median. 
    \label{fig:box_tmqi}}
\end{figure}

Figure \ref{fig:box_tmqi} shows box-plots of TMQI scores for 140 fused images.
Similarly to Fig.\label{fig:box_dh}, the difference among Figs.\ref{subfig:tmqi_case4}, \ref{subfig:tmqi_case5}, and \ref{subfig:tmqi_case6} are exposure values $\{v_i\}$ that input multi-exposure images have.
From Fig.\ref{subfig:tmqi_case4}, when clear input multi-exposure images are given, all three MEF methods had high score.
However, from Figs.\ref{subfig:tmqi_case5} and \ref{subfig:tmqi_case6}, when unclear input images are given, Mertens's method cannot guarantee the high quality of resulting images.
In contrast, the proposed scheme can maintain the high quality of resulting images by using SSLA, even if unclear input images are given.
Besides, from Figs.\ref{subfig:tmqi_case4} and \ref{subfig:tmqi_case6},
the minimum score of the proposed scheme is lower than that of SSLA
because input multi-exposure images contained many saturated white pixels.
From Fig.\ref{subfig:tmqi_case5}, as well as the result of hue difference (See Fig.\ref{fig:box_dh1}),
the proposed method can maintain the performance of SSLA by removing some bright input images.

\begin{figure}[t]
    \centering
    \subfigure[Conventional MEF \cite{Mertens}]{
		\label{subfig:ldr1}
		\includegraphics[width = 0.48\columnwidth]{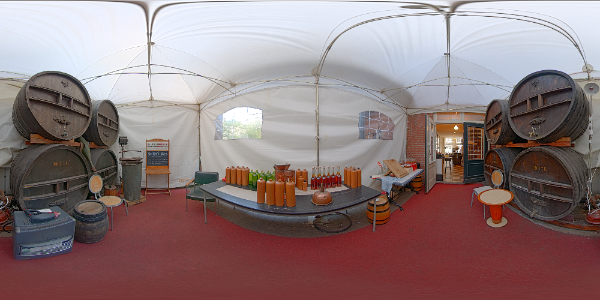}}
	\subfigure[Tone-mapping \cite{Fattal}]{
		\label{subfig:tm1}
		\includegraphics[width = 0.48\columnwidth]{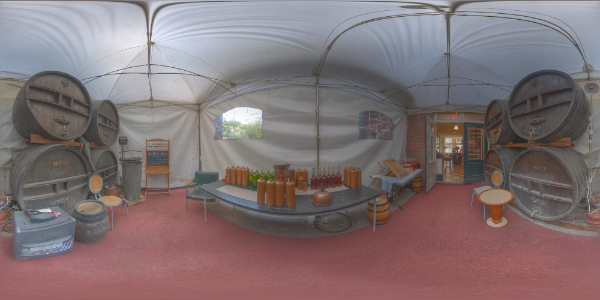}}
	\subfigure[Only SSLA]{
		\label{subfig:ssla1}
		\includegraphics[width = 0.48\columnwidth]{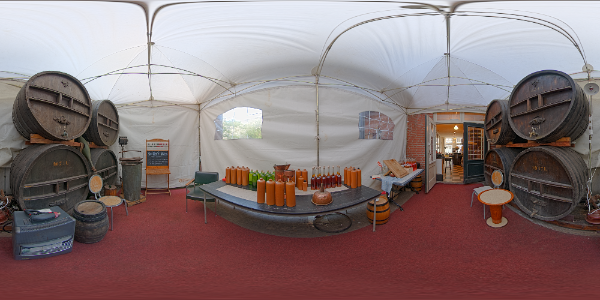}}
	\subfigure[Proposed]{
		\label{subfig:proposed1}
		\includegraphics[width = 0.48\columnwidth]{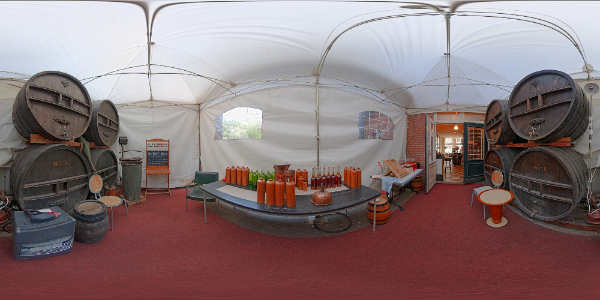}}
    \caption{Example of resulting images\\ ($\{v_i\}=\{-4,-2,0,2,4\}$[EV]).}
    \label{fig:result1}
\end{figure}

\begin{figure}[t]
    \centering
    \subfigure[Conventional MEF \cite{Mertens}]{
		\label{subfig:ldr2}
		\includegraphics[width = 0.48\columnwidth]{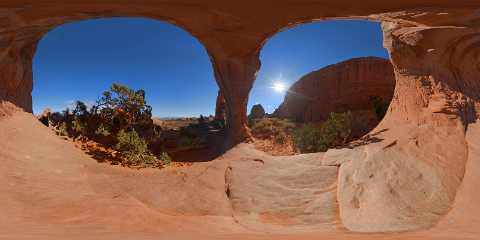}}
	\subfigure[Tone-mapping \cite{Fattal}]{
		\label{subfig:tm2}
		\includegraphics[width = 0.48\columnwidth]{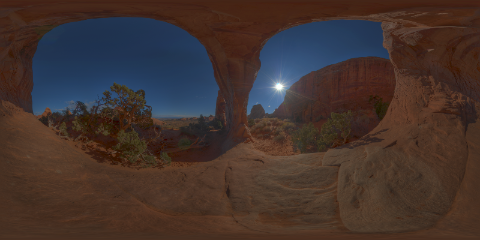}}
	\subfigure[Only SSLA]{
		\label{subfig:ssla2}
		\includegraphics[width = 0.48\columnwidth]{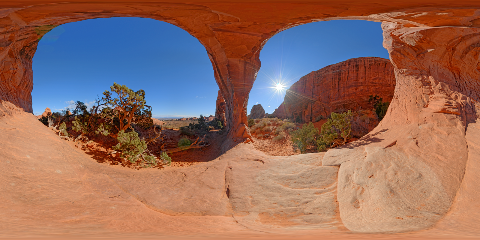}}
	\subfigure[Proposed]{
		\label{subfig:proposed2}
		\includegraphics[width = 0.48\columnwidth]{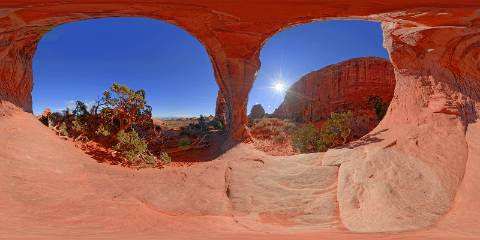}}
    \caption{Example of resulting images ($\{v_i\}=\{-4,-2,0\}$[EV]).}
    \label{fig:result2}
\end{figure}

Figure \ref{fig:result1} and \ref{fig:result2} show examples of resulting images.
From Figs.\ref{fig:result1} and \ref{fig:result2},
compared with images generated by Conventional MEF (a) and Tone-mapping (b),
images produced by the proposed scheme clearly represent the scenes, especially in terms of the shadow areas.
Also, compared with SSLA (c), the proposed scheme improved the color representation.
Therefore, the proposed scheme using SSLA and hue correction is effective for improving the quality of images fused by a MEF method.

\section{Conclusion}

In this paper, we propose a novel hue-correction scheme for MEF.
Hue correction in the proposed scheme is performed by replacing the maximally saturated colors of a fused image with those of an HDR one.
The HDR image is generated from input multi-exposure images, by removing the non-linearity of camera response $f$.
In addition, the use of SSLA in the proposed scheme enables us to generate higher-quality images than conventional MEF methods, regardless of the exposure condition of input images.

Experimental results showed that the high estimation accuracy of the function $f$ improve the performance of hue correction.
Experimental results showed the effectiveness of the proposed scheme in terms of the hue difference $\Delta H$ and TMQI under various exposure conditions.

\bibliographystyle{ieicetr}
\bibliography{refs}


\profile[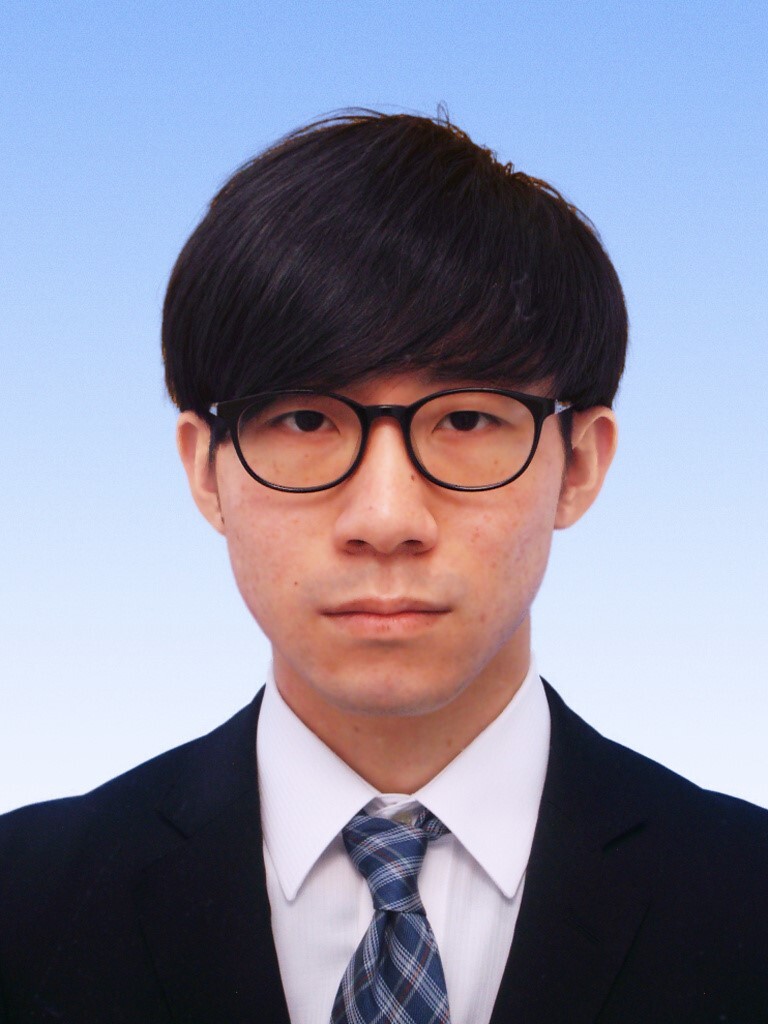]{Kouki Seo}{recieved his B.Eng degree from Tokyo Metropolitan University, Japan, in 2019. From 2019, he has been a Master course student at Tokyo Metropolitan University. His research interests include image processing.}

\profile[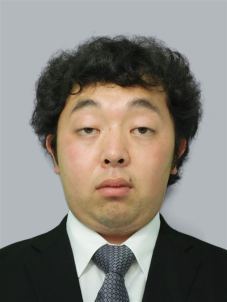]{Chihiro Go}{recieved his B.Eng degree from Tokyo University of Agriculture and Technology, Japan, in 2017. He is a Master course student at Tokyo Metropolitan University, Japan. His research interests include image processing.}

\profile[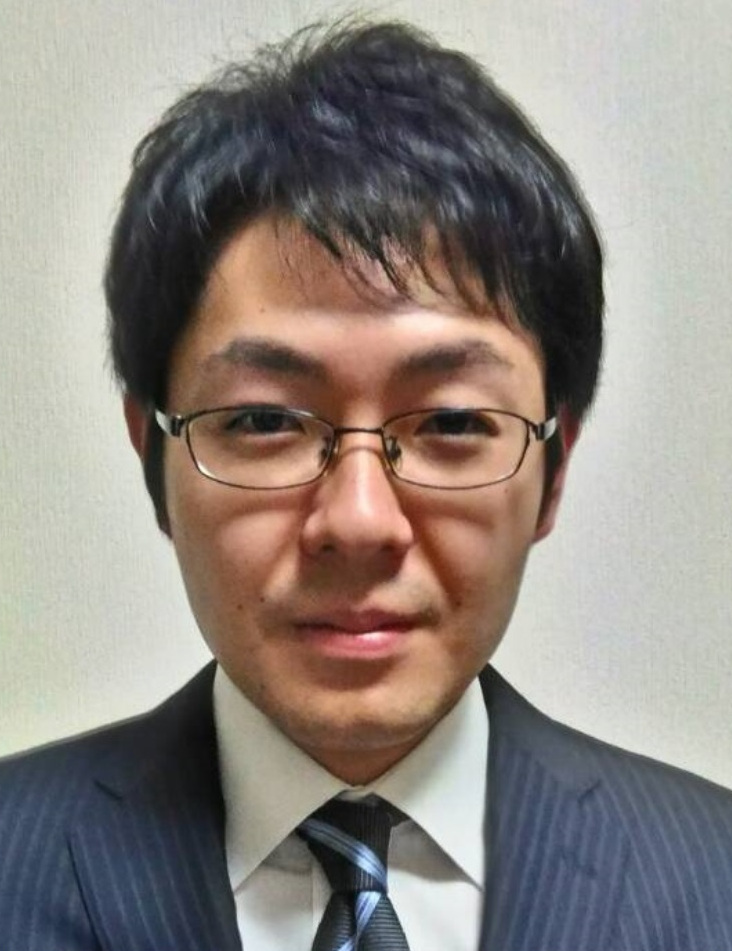]{Yuma Kinoshita}{received the B.Eng. and M.Eng. degrees from Tokyo Metropolitan University, Japan, in 2016 and 2018, respectively, where he is currently pursuing the Ph.D. degree. His research interest is in the area of image processing. He is a Student Member of APSIPA and IEICE. He received the IEEE ISPACS Best Paper Award, in 2016, the IEEE Signal Processing Society Japan Student Conference Paper Award, in 2018, the IEEE Signal Processing Society Tokyo Joint Chapter Student Award, in 2018, and the IEEE GCCE Excellent Paper Award (Gold Prize), in 2019.}

\profile[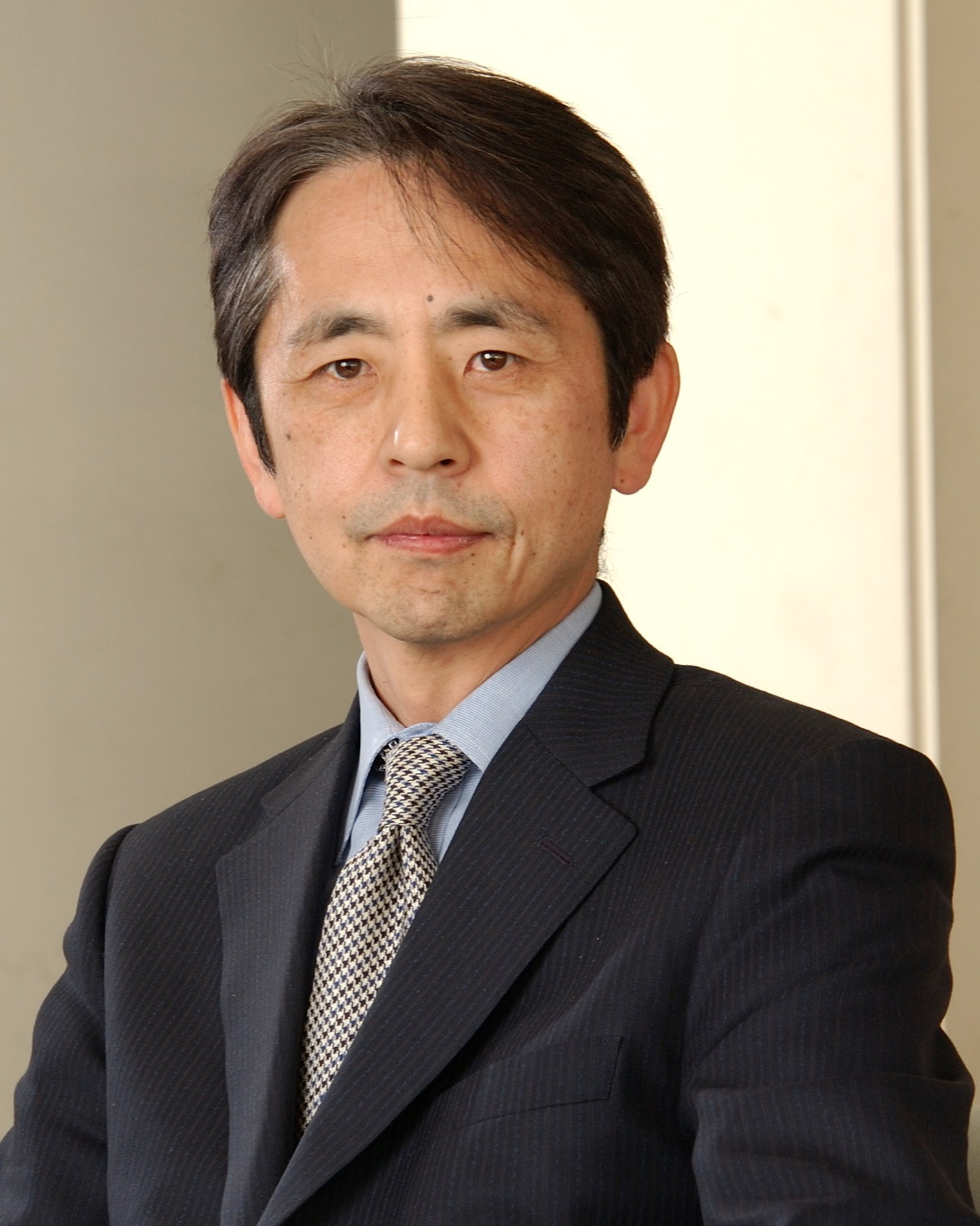]{Hitoshi Kiya}{received the B.E. and M.E. degrees from the Nagaoka University of Technology, in 1980 and 1982, respectively, and the Dr.Eng. degree from Tokyo Metropolitan University, in 1987. In 1982, he joined Tokyo Metropolitan University, where he became a Full Professor, in 2000. From 1995 to 1996,
he was a Visiting Fellow with The University of Sydney, Sydney, NSW, Australia. He served as the Inaugural Vice President (Technical Activities) of APSIPA, from 2009 to 2013, and the Regional Director-at-Large for Region ten of the IEEE Signal Processing Society, from 2016 to 2017. He was also the President of the IEICE Engineering Sciences Society, from 2011 to 2012. He currently serves as the President of APSIPA. He is a Fellow of IEICE and ITE. He served as the Vice President and the Editor-in-Chief for IEICE Society Magazine and Society Publications. He was a recipient of numerous awards, including nine best paper awards. He was the Chair of two technical committees and a member of nine technical committees, including the APSIPA Image, Video, and Multimedia Technical Committee (TC) and the IEEE Information Forensics and Security TC. He has organized a lot of international conferences in such roles as the TPC Chair of the IEEE ICASSP 2012 and the General Co-Chair of the IEEE ISCAS 2019. He was an Editorial Board Member of eight journals, including the IEEE TRANSACTIONS ON SIGNAL PROCESSING, Image Processing, and Information Forensics and Security.}

\end{document}